\documentclass[runningheads, twocolumn]{llncs}

\usepackage[a4paper,left=20mm,right=20mm,top=40mm,bottom=40mm]{geometry}
\usepackage[T1]{fontenc}
\usepackage{cite}
\usepackage{amsmath,amssymb,amsfonts}
\usepackage{algorithmic}
\usepackage{graphicx}
\usepackage{textcomp}
\usepackage{xcolor}
\usepackage{multirow}
\usepackage[hyphens]{url}
\usepackage{caption}
\usepackage{subcaption}
\usepackage{booktabs}
\usepackage{tabularx}
\usepackage{enumitem}
\usepackage[colorlinks=true, linkcolor=blue, citecolor=blue, urlcolor=black]{hyperref}
\usepackage[capitalize, nameinlink]{cleveref}
\crefname{figure}{Fig.}{Figs.}
\Crefname{figure}{Fig.}{Figs.}

\usepackage{amssymb}
\usepackage{multirow}
\usepackage{makecell}

\usepackage{xspace}
\newcommand{\BfPara}[1]{{\noindent\bf#1.}\xspace}
\newcommand{\AID}[1]{{\setlength{\fboxsep}{1.5pt}\ensuremath{\boxed{\textbf{#1}}}}\hspace{0.2em}}

\usepackage{etoolbox}
\makeatletter
\apptocmd{\@maketitle}
  {\begin{abstract}

Vehicle diagnostics has become essential for detecting in-vehicle errors and ensuring safety. While the Unified Diagnostic Services (UDS) protocol is widely adopted for diagnostic operations, it relies on the ISO 15765-2 standard as the transport protocol over the Controller Area Network (CAN), which was designed without inherent security considerations. In this paper, we identify eight novel attack scenarios that exploit specific transport layer mechanisms in the ISO 15765-2 standard, including Flow Control manipulation, Sequence Number violations, and error handling abuses. We evaluate these attacks on a real passenger vehicle using two distinct diagnostic tools to demonstrate their practical impact. Our results confirm that three of these attack scenarios successfully induce denial of diagnostic services, leading to abnormal diagnostic results such as concealed faults and manipulated sensor readings. These findings highlight critical vulnerabilities that can deceive technicians and drivers, potentially exposing vehicles to significant safety risks.


\end{abstract}\vspace{1em}}
  {}
  {\typeout{Failed to patch @maketitle}}
\makeatother

\begin{document}
\title{The Vehicle May Be Sick: Denial of Diagnostic Services by Exploiting the CAN Transport Protocol}

\titlerunning{The Vehicle May Be Sick}

\author{Seungjin Baek\inst{1} \and Seonghoon Jeong\inst{2} \and Huy Kang Kim\inst{1}}
\institute{School of Cybersecurity, Korea University, Seoul, Republic of Korea\\
\email{baekseungjin@korea.ac.kr}, \email{cenda@korea.ac.kr} \and
Division of Artificial Intelligence Engineering, Sookmyung Women's University, Seoul, Republic of Korea\\
\email{seonghoon@sookmyung.ac.kr}}

\authorrunning{Baek et al.}

\maketitle
\section{Introduction} \label{sec:introduction}

As modern vehicles integrate an increasing number of technologies and functions, in-vehicle complexity continues to grow, which can lead to potential errors and defects~\cite{C1, C2}.
Vehicle diagnostics has therefore become essential as a technology for detecting errors and defects within internal vehicle systems and for enabling smooth maintenance, management, and testing throughout the vehicle life cycle.
Diagnostic services provide direct access to Electronic Control Units (ECUs) through standardized protocols such as Unified Diagnostic Services (UDS)~\cite{B0}, allowing technicians to read fault codes, monitor sensor data, and reprogram vehicle systems.

However, because diagnostic interfaces provide direct access to internal vehicle systems, they present an attractive attack surface for adversaries~\cite{E4, E5}.
If vehicle diagnostic functions are exploited or compromised, the consequences can severely impact vehicle safety and reliability.
Prior research has confirmed vulnerabilities in vehicle diagnostic services that can threaten occupant safety during driving, with exploitation methods focusing primarily on the application layer~\cite{I1, B5, R3}.
For instance, attackers who gain access to the Controller Area Network (CAN) bus through compromised infotainment systems~\cite{E1}, ECU vulnerabilities~\cite{E2, E3}, or wireless interface flaws~\cite{E4} can inject malicious messages that disrupt diagnostic operations, potentially masking critical fault conditions or preventing necessary maintenance.

While existing research has primarily focused on application-layer attacks against diagnostic protocols~\cite{I0}, the transport layer remains largely unexplored from a security perspective.
The ISO 15765-2 standard~\cite{B1}, commonly known as CAN Transport Protocol (CAN-TP) or ISO-TP, defines the transport protocol for diagnostic communication over CAN, yet it was designed without considering security implications.
This presents a significant research gap, as the transport layer handles message segmentation, flow control, and reassembly---mechanisms that, if exploited, can fundamentally disrupt diagnostic communication regardless of application-layer protections.
Furthermore, the widespread adoption of the ISO 15765-2 standard as the mandatory transport protocol for emissions-related On-Board Diagnostics (OBD)~\cite{B2,B3} and OBDonUDS~\cite{B4} means that vulnerabilities at this layer could affect vehicles across multiple manufacturers.

This paper addresses this research gap by focusing on the transport layer and deriving eight novel attacks that exploit frame characteristics, transmission mechanisms, and error handling in the ISO 15765-2 standard.
We analyze the impact of these attacks on actual vehicle diagnostic processes, demonstrating that specific exploitations can induce denial of diagnostic services that prevents proper vehicle diagnosis.
Through comprehensive experiments targeting a 2021 Hyundai Elantra CN7 passenger vehicle with two distinct diagnostic tools, we demonstrate that three of these eight attack scenarios successfully impact the real-world diagnostic process. The contributions of this paper can be summarized as follows:

\begin{itemize}
\item This research is the first to derive eight novel attacks based on the ISO 15765-2 standard and demonstrate their impact on diagnostic communication in a real passenger vehicle.
\item We identify fundamental design flaws in the CAN transport protocol resulting from a lack of security considerations, and present countermeasures for each attack.
\item We provide motivation for vehicle Original Equipment Manufacturers (OEMs) to verify whether their vehicles are vulnerable to the proposed attacks and whether adequate mitigations exist, given the widespread adoption of the ISO 15765-2 standard.
\end{itemize}

The remainder of this paper is organized as follows: Section~\ref{sec:background} provides background on the ISO 15765-2 standard and related works. Section~\ref{sec:adversary} describes the adversary model and prerequisites. Section~\ref{sec:attack} presents the eight attack scenarios derived from the ISO 15765-2 standard. Section~\ref{sec:experiment} describes the experimental setup and results. Finally, Section~\ref{sec:discussion} discusses the implications of our findings and future work.
\section{Background} \label{sec:background}

\subsection{ISO 15765-2: Diagnostic Communication over CAN}
\begin{figure}[t]
    \centering
    \includegraphics[width=\linewidth]{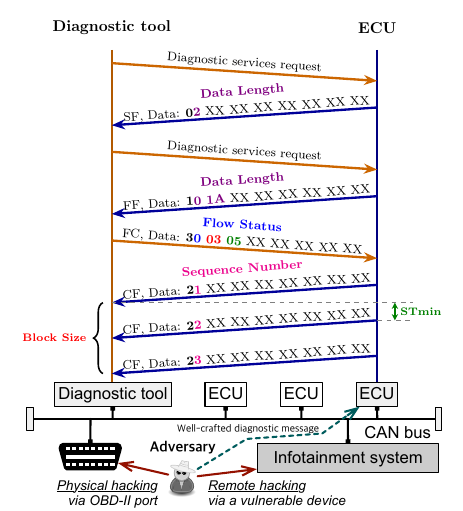}
    \caption{Benign transmission sequence and the adversary model}
    \label{fig:sequence}
\end{figure}

\begin{figure*}[t]
    \centering
    \includegraphics[width=\linewidth]{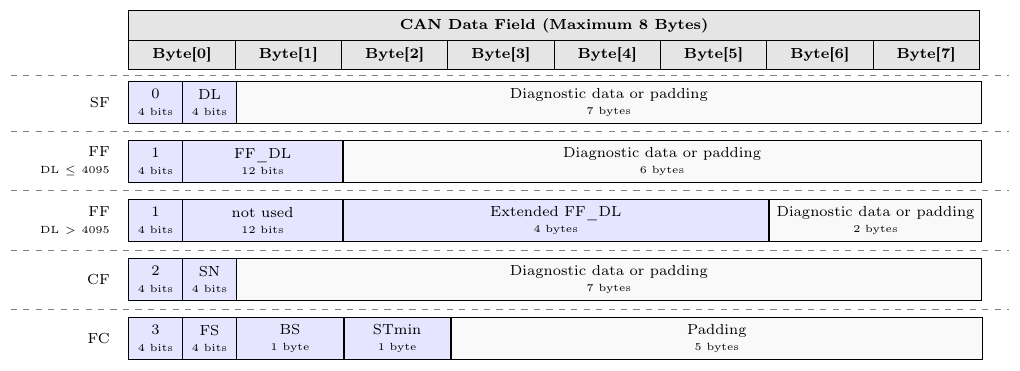}
    \caption{Data field structures for SF, FF, CF, and FC frames}
    \label{fig:bytefield}
\end{figure*}

The ISO 15765-2 standard defines the transport protocol and network layer for diagnostic communication over in-vehicle CAN.
This transport protocol enables the transmission of large diagnostic data by segmenting it into multiple fragments and reconstructing the original data through sequential reassembly.
The protocol is designed to overcome the data size limitations inherent in CAN messages, coordinate sender-receiver interactions through flow control mechanisms, and minimize data loss during transmission.
\autoref{fig:bytefield} illustrates the data field formats of CAN messages exchanged between sender and receiver during diagnostic communication.

\BfPara{Single Frame (SF)} This frame is used when the diagnostic data fits within a single CAN message, eliminating the need for fragmentation and flow control.
The first byte contains a 4-bit Data Length (DL) field indicating the length of the diagnostic data payload.
The receiver ignores frames where SF\_DL is 0.

\BfPara{First Frame (FF)} This frame signals the start of a multi-frame transmission when diagnostic data exceeds the capacity of a single CAN message, and includes a 12-bit DL field.
When the diagnostic data size exceeds the maximum FF\_DL value of 4,095, an escape sequence extends the data length specification to 4 bytes, supporting transmissions of up to approximately 4~GB.
The receiver ignores frames where FF\_DL is 7 or less, or where FF\_DL does not match the actual diagnostic data size.

\BfPara{Consecutive Frame (CF)} This frame carries the remaining diagnostic data in sequence according to flow control parameters and includes a 4-bit Sequence Number (SN).
The CF\_SN increments by 1 for each consecutively transmitted CF and wraps around to 0 upon reaching the maximum value.
Each CF must maximize its diagnostic data payload; only the final CF in a transmission may contain padding.

\BfPara{Flow Control (FC)} Upon receiving an FF, the receiver uses this frame to regulate the flow of subsequent CFs based on its reception capability.
This frame contains a 4-bit Flow Status (FS), a 1-byte Block Size (BS), and a 1-byte Separation Time minimum (STmin).
The FS field defines four possible values: FS\_ContinueToSend (CTS) to proceed with transmission, FS\_Wait to pause reception for a system-defined duration, FS\_Overflow to indicate that the received FF\_DL exceeds the buffer capacity, and FS\_Reserved for undefined values.
Notably, FS\_Wait may be transmitted consecutively up to Wait Frame Transmission max (WFTmax) times; FS\_Overflow is valid only in the first FC following an FF; and FS\_Reserved terminates message transmission as it represents an invalid state.
FC\_BS specifies the number of CFs that can be received before requiring another FC; a value of 0 indicates that all remaining frames should be transmitted without additional flow control.
FC\_STmin defines the minimum time interval between consecutive CFs, supporting intervals up to 127~ms.
When FS is set to Wait or Overflow, the FC\_BS and FC\_STmin fields are ignored.

All frames must use system-configured padding; minimizing padding can reduce bus load.
\autoref{fig:sequence} illustrates the frame transmission process during diagnostic communication. The receiver ignores any frames transmitted before receiving an initial SF or FF.
When full-duplex communication is supported, endpoints can simultaneously transmit and receive with different peers.
If an SF or FF is received during an ongoing reception, the current session is terminated and a new reception session begins.
A timeout occurs if the receiver does not receive the expected frame at the correct time or in the proper sequence.

\subsection{Related Works}
Kurachi \textit{et~al.}~\cite{R2} demonstrated a Man-in-the-Middle attack vulnerability where, following successful authentication between an ECU and a diagnostic tool, an attacker intercepts the communication, blocks messages from the diagnostic tool, and directly transmits malicious service requests that the ECU accepts as valid commands.
Nie \textit{et al.}~\cite{I2} discovered seed-and-key vulnerabilities in the UDS Security Access authentication mechanism that enabled unauthorized ECU access. Similarly, Kiley~\cite{I3} identified Security Access vulnerabilities in in-vehicle modules and demonstrated that exploiting these flaws could enable unauthorized battery management system firmware uploads.
More recent research has revealed that diagnostic services lacking authentication requirements can also be exploited.
Chatterjee \textit{et al.}~\cite{R3} investigated the security impact on medium and heavy-duty vehicles through exploitation of UDS diagnostic services and ISO-TP. Their work demonstrated that attackers can interrupt periodic ECU message transmissions and induce denial of both diagnostic sessions and diagnostic communication.
Han \textit{et al.}~\cite{I1} further demonstrated that permitted diagnostic services in certain vehicles can trigger sudden vehicle movements or stops during driving, directly threatening occupant safety.
Ren \textit{et al.}~\cite{B5} similarly revealed that UDS diagnostic service injection and session attacks can reboot ECUs or cause denial of steering and throttle control while driving.

Regarding defensive measures, Yekta \textit{et al.}~\cite{R4} proposed a methodology for detecting sophisticated UDS attacks within a Vehicle Security Operations Center.
Their approach, combining logging, contextual log data, and detection mechanisms, achieved higher attack detection coverage for known UDS attacks compared to existing AUTOSAR security events.
Weiss \textit{et~al.}~\cite{R1} proposed a methodology for automatically detecting endpoints in in-vehicle CAN communication using ISO-TP. By addressing limitations of existing application-layer-based endpoint detection, they demonstrated improved detection reliability in real vehicles, reducing potential attack surfaces and highlighting implications for future penetration testing and security analysis.
Researchers have also investigated methods to provide confidentiality, integrity, and authentication for CAN message transmission.
Yushev \textit{et al.}~\cite{R5} designed, implemented, and evaluated a secure communication channel for applying transport layer security to the CAN bus.
They demonstrated feasibility for low time-sensitivity applications through optimization leveraging the ISO-TP segmentation and reassembly mechanism.

The research surveyed above primarily focuses on application-layer vulnerabilities and their exploitation, with countermeasure research following similar trends.
This application-layer focus presents a limitation: the attacks proposed in this paper operate at the transport layer and therefore remain unaffected by such countermeasures.
Accordingly, this paper proposes eight novel attacks exploiting the frame characteristics, transmission mechanisms, and error handling defined in the ISO 15765-2 standard, thereby addressing this research gap and demonstrating their impact on the vehicle diagnostic process.
\section{Adversary Model} \label{sec:adversary}

\BfPara{Prerequisites}
We assume the attacker has direct access to the target CAN bus, enabling both passive monitoring of transmitted CAN messages and active injection of arbitrary CAN messages.
Such access can be achieved through the methods discussed in Section~\ref{sec:introduction}, including compromised infotainment systems, ECU vulnerabilities, and wireless interface exploits. Additionally, attackers can leverage vulnerable OBD dongles~\cite{A0, A1} to gain the required CAN bus access.

\BfPara{Knowledge}
The attacker must identify the CAN identifier range used for diagnostic communication within the target vehicle.
For emissions-related diagnostic services, the ISO 15765-4 standard~\cite{A2} legally mandates a specific addressing scheme.
This standard requires diagnostic tools to use 0x7DF for functional addressing (broadcast requests to multiple ECUs), 0x7E0--0x7E7 for physical addressing (requests to specific ECUs), and 0x7E8--0x7EF for ECU responses.
For UDS-based enhanced diagnostic services, 0x7DF is commonly used for broadcast requests, while other addressing follows OEM-specific definitions, requiring CAN bus monitoring for identification.
Through such monitoring, the attacker can observe that the arbitration ID range 0x700--0x7FF is typically used for diagnostic communication.
This design choice assigns diagnostic identifiers a lower arbitration priority than driving-related identifiers, preventing interference with safety-critical vehicle functions.

\BfPara{Goal}
The attacker's objective is to induce denial of diagnostic services by exploiting vulnerabilities in the ISO 15765-2 transport protocol implementation.
Successful attacks prevent diagnostic sessions from completing normally or cause them to produce abnormal results, effectively disrupting vehicle maintenance and fault detection capabilities.
\section{Attack Scenarios} \label{sec:attack}

\begin{table*}[t]
\centering
\caption{Systematic classification of proposed ISO 15765-2 transport layer attacks categorized by frame types, exploited protocol mechanisms, and relevant standard clauses.}
\label{tab:attack_analysis_final}
\vspace{5pt}
\renewcommand{\arraystretch}{0.9}
\small
\begin{tabularx}{\textwidth}{>{\raggedright\arraybackslash}X *{4}{>{\centering\arraybackslash}p{0.9cm}} >{\centering\arraybackslash}p{2.8cm} >{\centering\arraybackslash}p{2.0cm} >{\centering\arraybackslash}p{2.1cm}}
    \toprule
    \multirow{2}{*}{\textbf{Attacks}} & \multicolumn{4}{c}{\textbf{Frame Characteristics}} & \textbf{Transmission} & \textbf{Error} & \textbf{ISO 15765-2} \\
    \cmidrule(lr){2-5}
    & \textbf{SF} & \textbf{FF} & \textbf{CF} & \textbf{FC} & \textbf{Mechanism} & \textbf{Handling} & \textbf{Clause} \\
    \midrule 
    \AID{A1}  Preceding FlowControl Attack     &            &            &            & \checkmark & \checkmark &            & \S 9.6.5.6 \\
    \AID{A2}  FlowStatus Wait Attack           &            &            &            & \checkmark &            & \checkmark & \S 9.6.5.1, \S 9.8.4 \\
    \AID{A3}  FlowStatus Overflow Attack       &            &            &            & \checkmark & \checkmark &            & \S 9.6.5.1 \\
    \AID{A4}  DataLength Attack                &            & \checkmark &            &            &            & \checkmark & \S 9.6.3.2 \\
    \AID{A5}  SequenceNumber Attack            &            &            & \checkmark &            &            & \checkmark & \S 9.6.4.4 \\
    \AID{A6}  Session Override Attack          & \checkmark & \checkmark &            &            & \checkmark &            & \S 9.8.3 \\
    \AID{A7}  Reserved Value Attack            &            &            &            & \checkmark &            & \checkmark & \S 9.6.5.2 \\
    \AID{A8}  Timeout Attack                   &            &            &            &            & \checkmark & \checkmark & \S 9.8.2 \\
    \bottomrule
\end{tabularx}
\end{table*}

\begin{figure*}[t!]
    \centering
        \begin{subfigure}[t]{0.329\textwidth} 
            \centering
            \includegraphics[width=\textwidth]{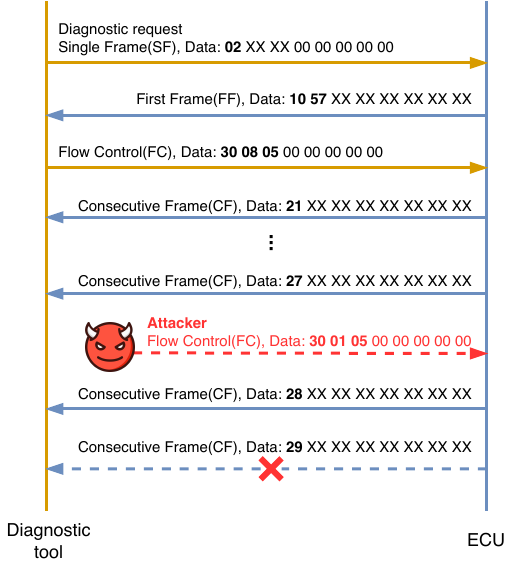}
            \caption{\AID{A1} Preceding FlowControl}
            \label{fig:Preceding_FC_Attack}
        \end{subfigure}
        \begin{subfigure}[t]{0.329\textwidth}
            \centering
            \includegraphics[width=\textwidth]{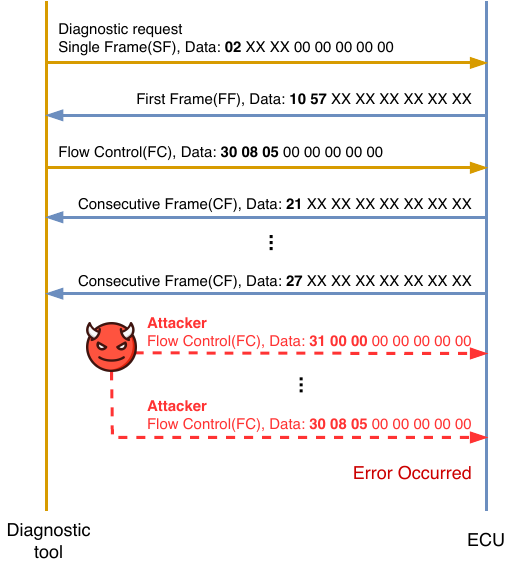}
            \caption{\AID{A2} FlowStatus Wait}
            \label{fig:FlowStatus_Wait_Attack}
        \end{subfigure}
        \begin{subfigure}[t]{0.329\textwidth}
            \centering
            \includegraphics[width=\textwidth]{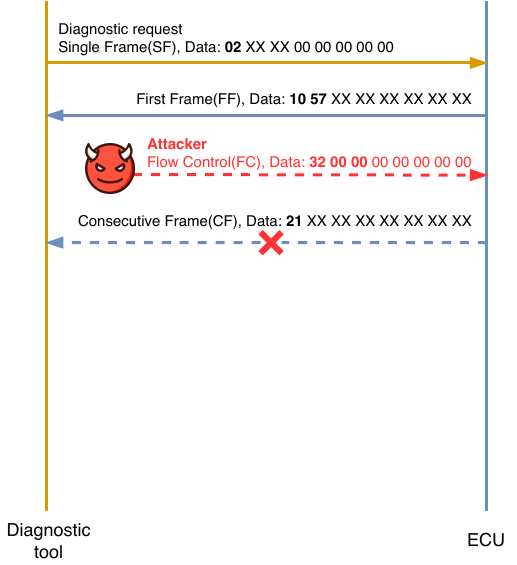}
            \caption{\AID{A3} FlowStatus Overflow}
            \label{fig:FlowStatus_Overflow_Attack}
        \end{subfigure}

        \vspace{1em}

        \begin{subfigure}[t]{0.329\textwidth} 
            \centering
            \includegraphics[width=\textwidth]{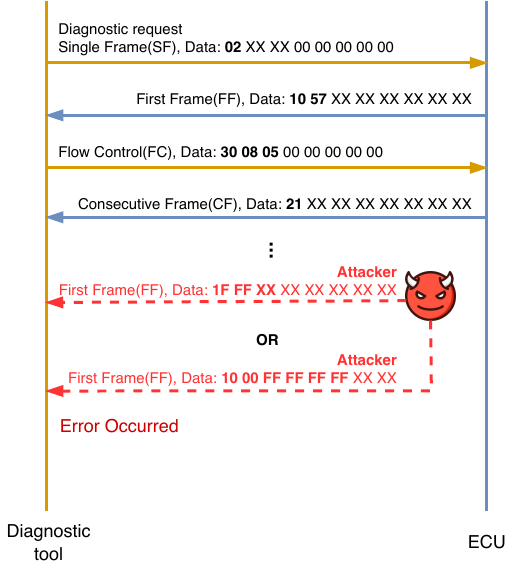}
            \caption{\AID{A4} DataLength}
            \label{fig:DataLength_Attack}
        \end{subfigure}
        \begin{subfigure}[t]{0.329\textwidth}
            \centering
            \includegraphics[width=\textwidth]{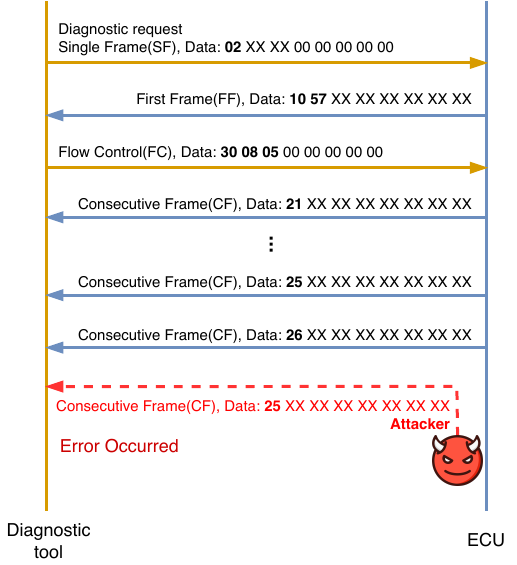}
            \caption{\AID{A5} SequenceNumber}
            \label{fig:SequenceNumber_Attack}
        \end{subfigure}
        \begin{subfigure}[t]{0.329\textwidth}
            \centering
            \includegraphics[width=\textwidth]{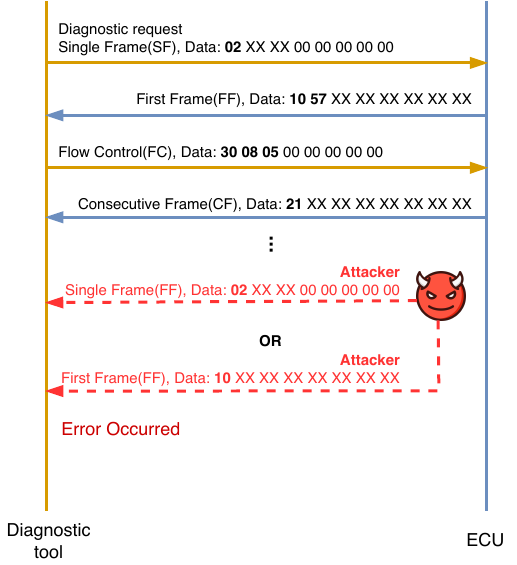}
            \caption{\AID{A6} Session Override}
            \label{fig:Receiving_mechanism_Attack}
        \end{subfigure}
        
        \vspace{1em}

        \begin{subfigure}[t]{0.329\textwidth} 
            \centering
            \includegraphics[width=\textwidth]{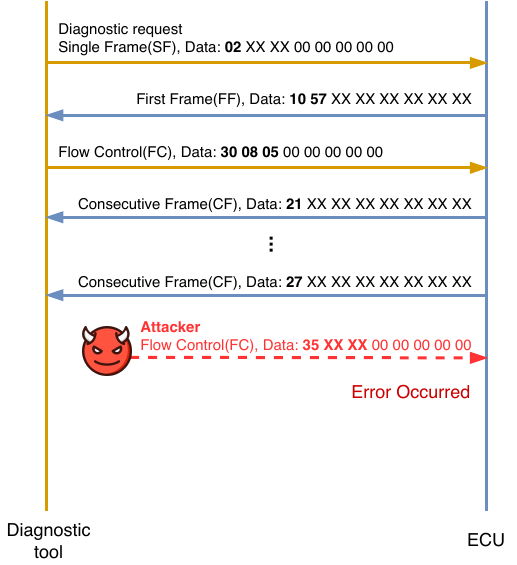}
            \caption{\AID{A7} Reserved Value}
            \label{fig:Reserved_value_Attack}
        \end{subfigure}
        \quad
        \begin{subfigure}[t]{0.329\textwidth}
            \centering
            \includegraphics[width=\textwidth]{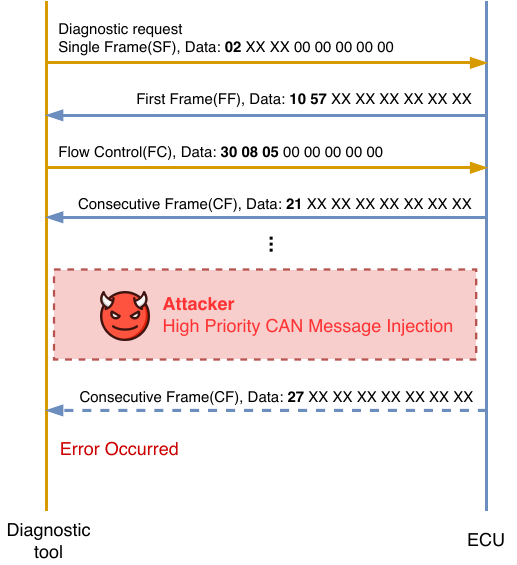}
            \caption{\AID{A8} Timeout}
            \label{fig:Timeout_Attack}
        \end{subfigure}
        
        \caption{Detailed flow of ISO 15765-2 standard based attack scenarios}
        \label{fig:Attack_Scenario}
\end{figure*}

This section details eight novel attack scenarios \AID{A1}--\AID{A8} from an attacker's perspective, illustrated in \autoref{fig:Attack_Scenario}. Each scenario is structured to first explain the relevant protocol mechanism, followed by the attacker's method of exploitation, and finally the resulting disruption to the diagnostic process.

\begin{figure*}[t]
    \centering
    \begin{subfigure}[t]{0.475\linewidth}
        \centering
        \includegraphics[width=\textwidth]{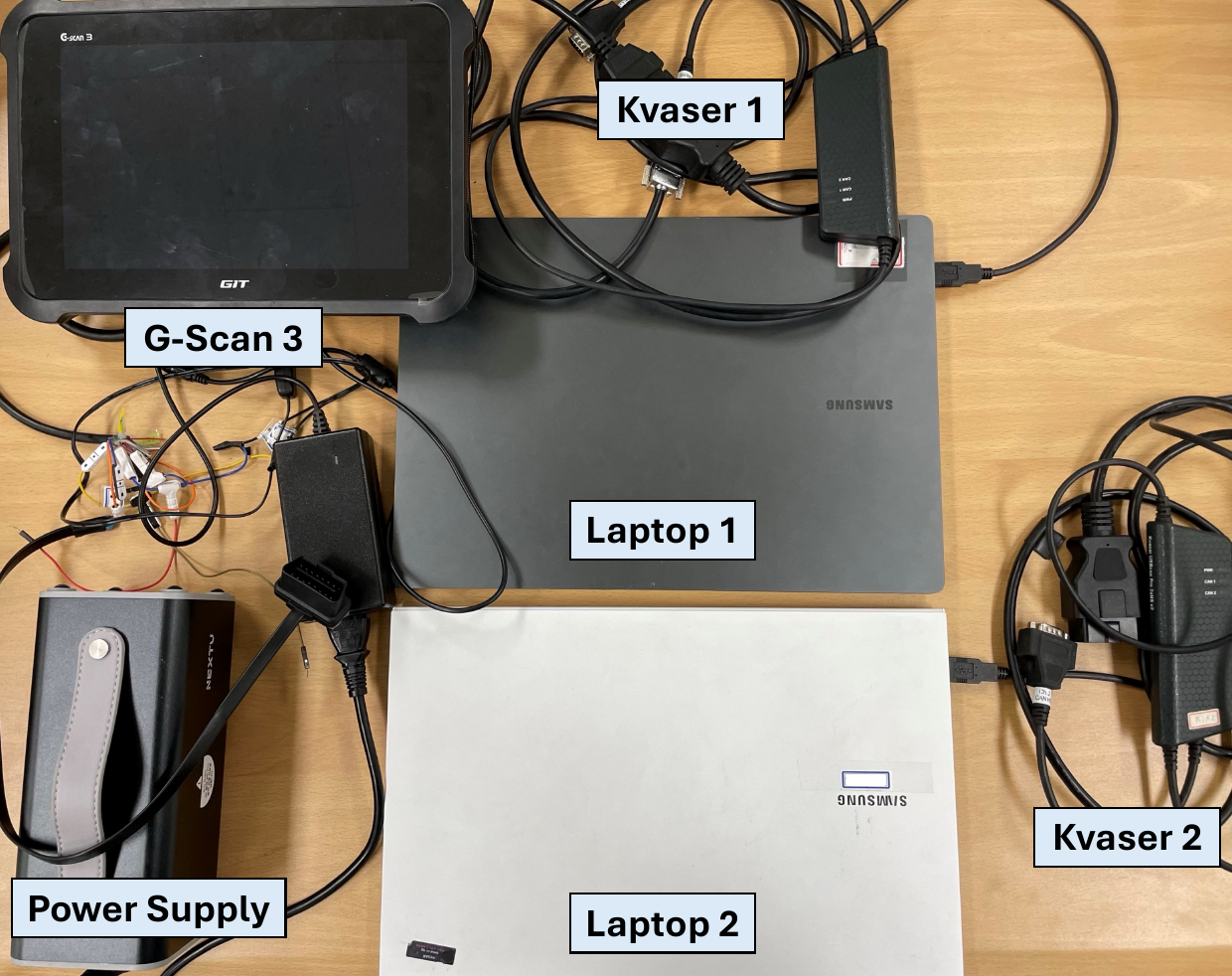}
        \caption{Using the Hyundai Motor Group OEM-level G-Scan3 diagnostic tool}
        \label{fig:gscan_testbed}
    \end{subfigure}
    \hfill
    \begin{subfigure}[t]{0.5\linewidth}
        \centering
        \includegraphics[width=\textwidth]{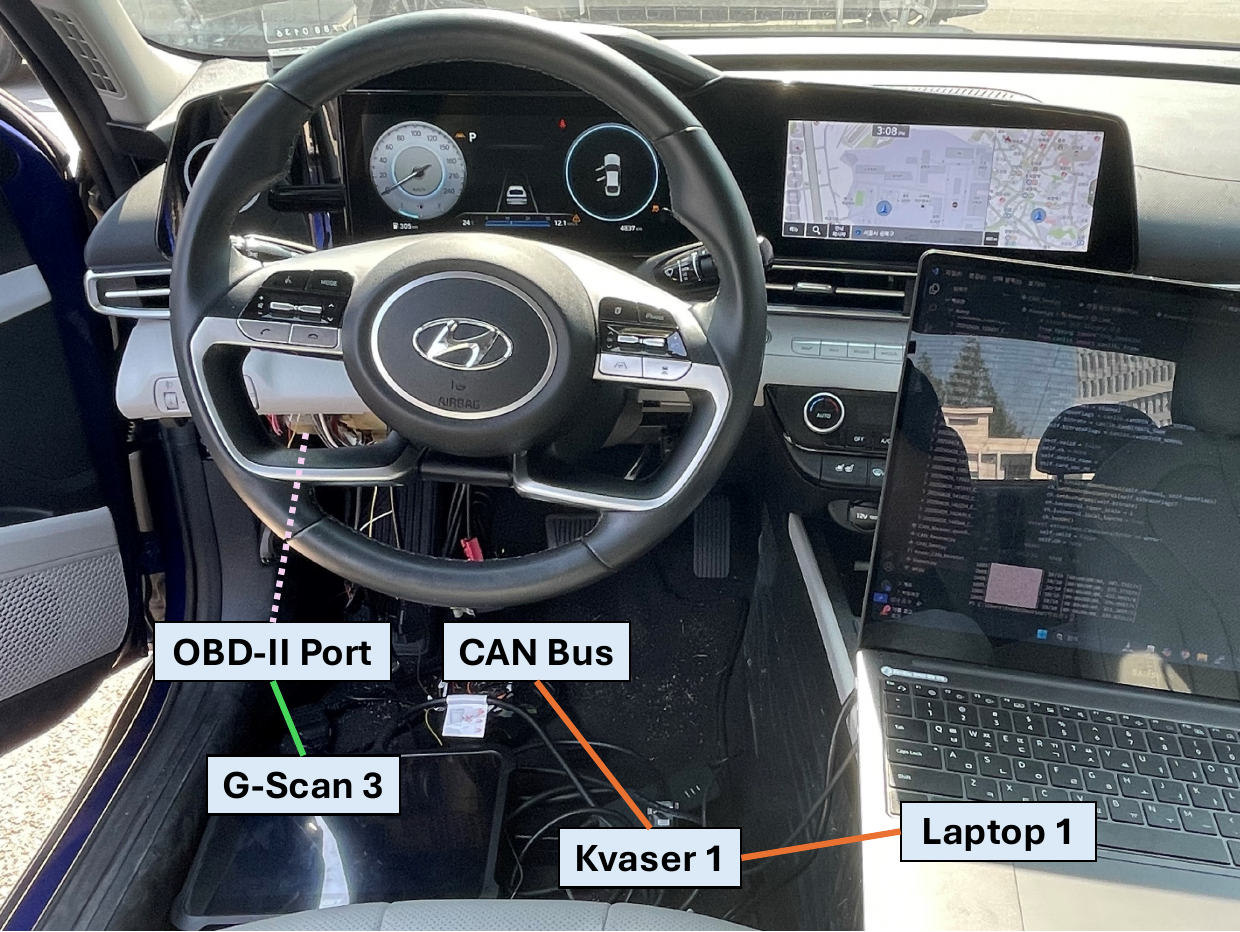}
        \caption{Configuration for practicing diagnostics and attacks in a 2021 Hyundai Elantra CN7 passenger vehicle}
        \label{fig:in_vehicle_setup}
    \end{subfigure}
    \caption{The setup also configures an OBDSCAN4.0 diagnostic tool based on ELM327 instead of G-Scan3. Laptop 1 performs the role of an attacker using Kvaser 1, and Laptop 2 monitors the CAN bus using Kvaser 2.}
    \label{fig:diagnostic_tool}
\end{figure*}

\begin{enumerate}[align=left]
\item[\AID{A1}] \textbf{Preceding FlowControl Attack:}
In diagnostic communication, the FC frame regulates the reception of CFs transmitted by the ECU.
By exploiting frame characteristics and transmission mechanisms, the attacker monitors CF reception and preemptively transmits a manipulated FC frame at the expected timing of the legitimate FC response, as shown in \autoref{fig:Preceding_FC_Attack}.
Since the protocol may accept parameters from a subsequent FC that differ from those of the preceding one, the attacker can set the Block Size (FC\_BS) to a minimum value, causing the omission of subsequent CFs.

\item[\AID{A2}] \textbf{FlowStatus Wait Attack:}
The Flow Status (FC\_FS) parameter manages CF reception. The system limits the maximum number of consecutive FC frames with FS\_Wait status using the WFTmax parameter.
If transmission resumes with an FS\_CTS frame after the maximum FS\_Wait count is reached but reception performance requirements are not met, the system treats this as an error and aborts reception.
As shown in \autoref{fig:FlowStatus_Wait_Attack}, the attacker exploits this by transmitting the maximum allowable number of FS\_Wait frames at the expected timing of the normal FC response.
Immediately following this, the attacker sends an FS\_CTS frame, causing the system to terminate message reception due to exceeding performance limits.

\item[\AID{A3}] \textbf{FlowStatus Overflow Attack:}
An FC frame with FS\_Overflow status is valid only when responding to the first received FF and causes immediate termination of message transmission.
As shown in \autoref{fig:FlowStatus_Overflow_Attack}, the attacker exploits this mechanism by injecting an FS\_Overflow FC frame immediately after detecting the first FF, thereby forcing the termination of the ongoing message transmission.

\item[\AID{A4}] \textbf{DataLength Attack:}
The FF\_DL field indicates the size of the diagnostic data payload. The receiver considers it an error if FF\_DL exceeds its available buffer size.
Exploiting this error handling mechanism (\autoref{fig:DataLength_Attack}), the attacker transmits an FF with FF\_DL set to the maximum 12-bit value (4,095) or the maximum 32-bit value (4,294,967,295) using the escape sequence, causing the receiver to abort message reception.

\item[\AID{A5}] \textbf{SequenceNumber Attack:}
The CF\_SN must increment sequentially by one within the range of 0x0 to 0xF. Any violation of this rule causes the receiver to abort reception.
As shown in \autoref{fig:SequenceNumber_Attack}, the attacker monitors CF reception and injects a CF with an arbitrary SN, deliberately violating the sequence rule and forcing the receiver to terminate the session.

\item[\AID{A6}] \textbf{Session Override Attack:}
The protocol dictates that if an SF or FF is received from the same CAN identifier during an active reception, the current reception is terminated and a new one is initiated.
The attacker exploits this mechanism (\autoref{fig:Receiving_mechanism_Attack}) by transmitting an arbitrary SF or FF via physical addressing to a specific ECU, thereby overriding and terminating the legitimate reception process.

\item[\AID{A7}] \textbf{Reserved Value Attack:}
Values in the range 0x3--0xF for the FC\_FS field are reserved. The transmitter treats any FC frame containing a reserved value as an error and aborts transmission.
As shown in \autoref{fig:Reserved_value_Attack}, the attacker monitors CF reception and transmits an FC frame with a reserved FC\_FS value at the expected timing of a normal FC response, causing immediate termination of the message transmission.

\item[\AID{A8}] \textbf{Timeout Attack:}
Timeouts are implemented during CF and FC reception to ensure system reliability in cases of frame loss, bus congestion, or delays.
While the system allows a tolerance margin (up to 50\% relative to the lower limit), a timeout event is treated as an error that terminates transmission.
The attacker exploits this (\autoref{fig:Timeout_Attack}) by flooding the bus with high-priority CAN messages, intentionally causing delays that trigger a timeout and abort the diagnostic session.
\end{enumerate}

Through an exhaustive analysis of logical flaws and edge cases in the ISO 15765-2 standard, we identified these non-heuristic attacks based on frame characteristics, transmission mechanisms, and error handling. \autoref{tab:attack_analysis_final} summarizes these findings.
\section{Experimental Evaluation} \label{sec:experiment}

\subsection{Diagnostic Tools}
The experimental setup follows the configuration in \autoref{fig:diagnostic_tool} to demonstrate the impact of attacks through the use of two diagnostic tools with distinct characteristics.
OBDSCAN4.0 is an affordable and easily accessible OBD-II aftermarket diagnostic tool that requests diagnostic services and displays results through compatible free diagnostic applications such as Car Scanner and Torque.
Specifically, it provides services to retrieve in-vehicle Diagnostic Trouble Codes (DTCs) and ECU information, monitor onboard sensors, and record data.
However, the tool does not support Secure Gateway (SGW) security authentication, restricting it to limited services. Furthermore, reliability issues may arise because the tool relies on an external CAN database to interpret results, rather than querying the vehicle directly.
The G-Scan3 (\autoref{fig:gscan_testbed}), developed by GIT (an affiliate of Hyundai Motor Group), supports OEM-level diagnostics for Hyundai/Kia vehicles. This tool performs SGW security authentication, enabling access to restricted diagnostic services such as variant coding, actuation tests, emission tests, and sensor calibration.
This ensures highly reliable diagnostic results for Hyundai/Kia vehicles. Although this experimental setup employs two diagnostic tools with distinct characteristics within a single vehicle, it can be configured with various OEM vehicles and diagnostic tools.

\begin{figure*}[t] 
    \centering
    \begin{subfigure}[b]{0.48\textwidth} 
        \centering
        \includegraphics[width=\textwidth]{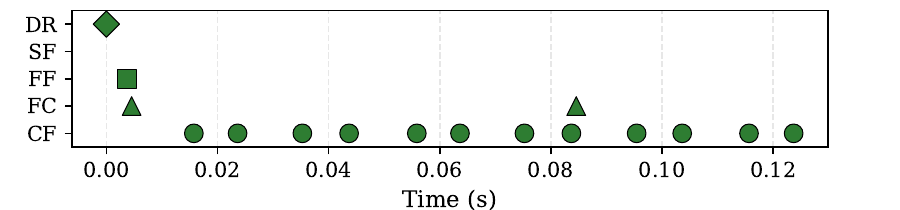}
        \caption{Normal}
        \label{fig:attack_normal}
    \end{subfigure}
    \hfill
    \begin{subfigure}[b]{0.48\textwidth}
        \centering
        \includegraphics[width=\textwidth]{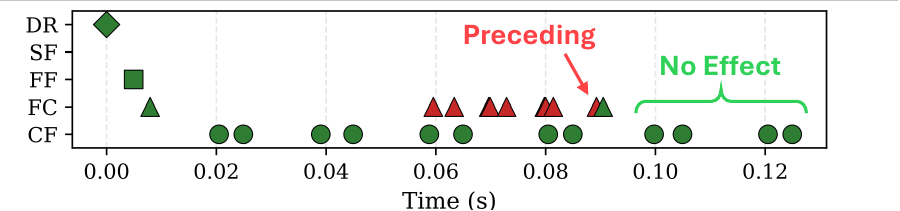}
        \caption{\AID{A1} Preceding FlowControl}
        \label{fig:attack_preceding_fc}
    \end{subfigure}
    
    \begin{subfigure}[b]{0.48\textwidth}
        \centering
        \includegraphics[width=\textwidth]{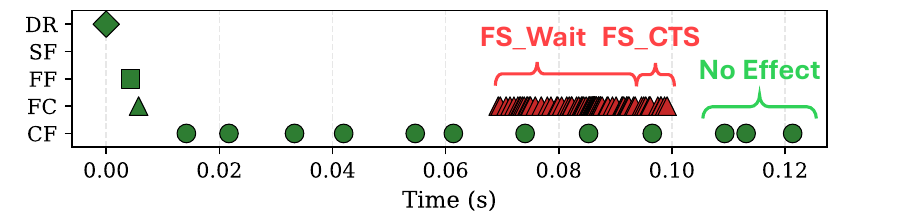}
        \caption{\AID{A2} FlowStatus Wait}
        \label{fig:attack_fs_wait}
    \end{subfigure}
    \hfill
    \begin{subfigure}[b]{0.48\textwidth}
        \centering
        \includegraphics[width=\textwidth]{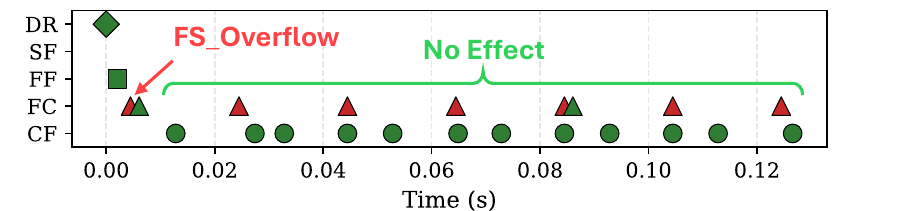}
        \caption{\AID{A3} FlowStatus Overflow}
        \label{fig:attack_fs_overflow}
    \end{subfigure}
    
    \begin{subfigure}[b]{0.48\textwidth}
        \centering
        \includegraphics[width=\textwidth]{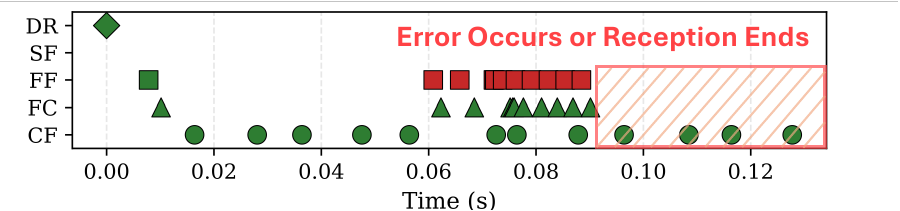}
        \caption{\AID{A4} DataLength}
        \label{fig:attack_ff_dl}
    \end{subfigure}
    \hfill
    \begin{subfigure}[b]{0.48\textwidth}
        \centering
        \includegraphics[width=\textwidth]{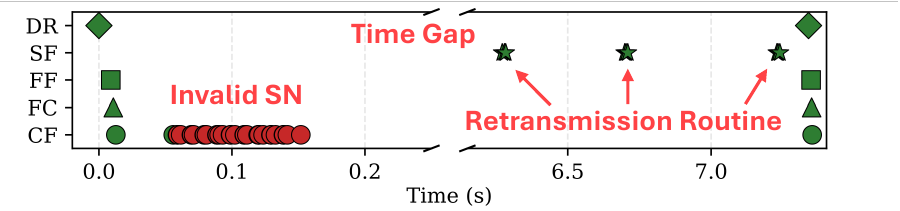}
        \caption{\AID{A5} SequenceNumber}
        \label{fig:attack_cf_sn}
    \end{subfigure}
    
    \begin{subfigure}[b]{0.48\textwidth}
        \centering
        \includegraphics[width=\textwidth]{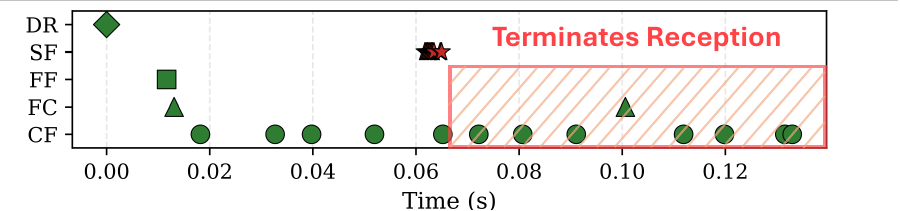}
        \caption{\AID{A6} Session Override (SF)}
        \label{fig:attack_receiving_sf}
    \end{subfigure}
    \hfill
    \begin{subfigure}[b]{0.48\textwidth}
        \centering
        \includegraphics[width=\textwidth]{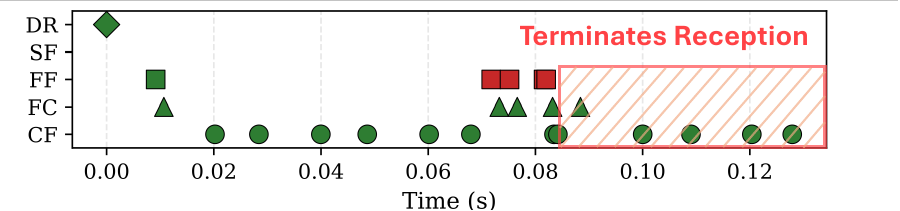}
        \caption{\AID{A6} Session Override (FF)}
        \label{fig:attack_receiving_ff}
    \end{subfigure}
    
    \begin{subfigure}[b]{0.48\textwidth}
        \centering
        \includegraphics[width=\textwidth]{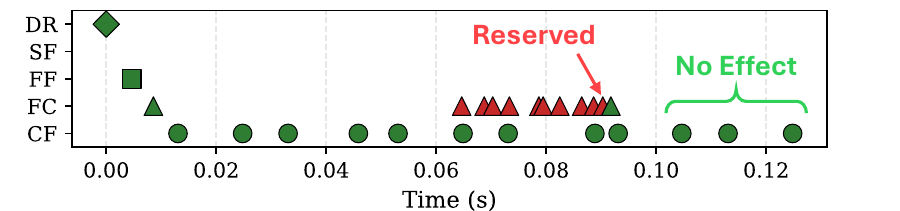}
        \caption{\AID{A7} Reserved Value}
        \label{fig:attack_reserved}
    \end{subfigure}
    \hfill
    \begin{subfigure}[b]{0.48\textwidth}
        \centering
        \includegraphics[width=\textwidth]{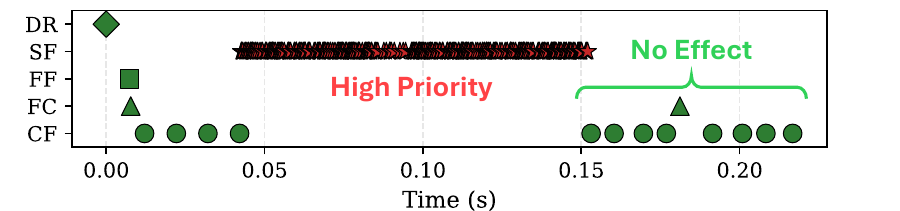}
        \caption{\AID{A8} Timeout}
        \label{fig:attack_timeout}
    \end{subfigure}
    \caption{CAN message traces for normal communication and each attack scenario. Green and red markers indicate normal and attack frames, respectively.}
    \label{fig:attack_traces}
\end{figure*}

\subsection{Attack Validation}
We execute each attack in the real vehicle diagnostic environment shown in {\autoref{fig:in_vehicle_setup}} and analyze the resulting responses at the diagnostic communication level. Notably, significant results were observed for only a subset of attacks.

\autoref{fig:attack_normal} shows the normal diagnostic communication process during the execution of the `ECU Identifiers' vehicle diagnostic function.
In \autoref{fig:attack_receiving_sf} and \autoref{fig:attack_receiving_ff}, the attacker sends an FF and an SF using physical addressing to induce the termination of the active reception session.
Because the attack does not generate an error, the CFs appeared to be transmitted normally afterward; however, abnormal diagnostic results were observed.
This indicates that the active reception session was interrupted by the initiation of a new reception session, overwriting the previous diagnostic data.
In \autoref{fig:attack_ff_dl}, the attacker sends an FF with the FF\_DL set to the maximum value, thereby exceeding the receiver's buffer size and inducing message reception termination.
Although message reception did not cease immediately, abnormal diagnostic results were observed.
This indicates that the induced error corrupted the results, or that a session-override behavior analogous to \AID{A6} was triggered.
In \autoref{fig:attack_cf_sn}, the attacker induces message reception termination by transmitting CFs that violate SN continuity.
Following the attack, CF transmission ceased immediately, and after a period of unresponsive delay, a diagnostic request retransmission routine occurred a fixed number of times to resume the diagnostic service. This indicates that the in-vehicle transport protocol implementation recognized the error and halted message reception, successfully inducing abnormal diagnostic results.
Other attacks demonstrated no influence on the diagnostic results, although they were executed as intended.
For \cref{fig:attack_preceding_fc,fig:attack_fs_wait,fig:attack_fs_overflow}, we can infer that the in-vehicle transport protocol implementation is designed to ignore or reject additionally transmitted FC frames.
Furthermore, for \cref{fig:attack_reserved,fig:attack_timeout}, we suspect that the implementation similarly ignores frames where the FC\_FS is set to a reserved value, or that the configured timeout margin is sufficiently wide.

\subsection[Attack Impact]{Attack Impact}
Focusing on the \AID{A4}--\AID{A6} attacks, which successfully induced abnormal diagnostic results, we explain their impact and potential risks to vehicle diagnostic functions using two distinct diagnostic tools.

\BfPara{ECU Identifiers}
The ECU identification function of the Car Scanner diagnostic application retrieves software versions, Vehicle Identification Numbers (VINs), and manufacturing and supply information for in-vehicle ECUs that are difficult to access directly.
Attacks \AID{A4}, \AID{A5}, and \AID{A6} cause diagnostic result omission by terminating the active reception session or interrupting message flow.
In particular, these attacks increase the diagnostic time by up to six times the normal duration. Furthermore, \AID{A6} induces `diagnostic data spraying' where a new reception session overwrites parts of the valid diagnostic results with initialization values.
These results render the identification of in-vehicle information and status through diagnostics impossible.

\BfPara{DTC Inquiry}
In-vehicle ECUs detect system errors and record them as DTCs to enable subsequent diagnosis.
Attacks \AID{A4}, \AID{A5}, and \AID{A6} interfere with this detection process and manipulate the count of queried DTCs.
Specifically, the G-Scan3 failed to receive DTC inquiry data normally; it displayed `0', a presumed default value, without indicating an error.
These results expose the vehicle to potential risks by concealing internal errors during the reporting process.

\BfPara{Functional Test}
Evaporative emission leak tests inspect the evaporative emission system for gas leaks to maintain vehicle performance and safety.
Although the G-Scan3 normally reports no leaks while the vehicle remains stationary, attacks \AID{A4}, \AID{A5}, and \AID{A6} manipulate these test results.
Specifically, the tool failed to receive valid leak test data and displayed ``gas leak detected''---likely a default state---without triggering an error code.
These results deceive in-vehicle systems, which can impact emissions-related inspections and lead to regulatory violations.

\BfPara{Variant Coding}
Variant coding functions allow for modifying specific ECU options or setting specification information. This capability adapts common hardware properties to specific vehicle configurations, activating necessary functions.
Attacks \AID{A4}, \AID{A5}, and \AID{A6} interfered with the transmission of coding request data and disrupted the normal operation of G-Scan3's coding functions.
These results can hinder the replacement of faulty vehicle components or restrict the activation of specific functions.
\section{Discussion and Future Work} \label{sec:discussion}

The eight novel attacks derived in this paper are based on the ISO 15765-2 standard, and vehicles adopting the Controller Area Network Flexible Data-Rate (CAN-FD) protocol may also be affected.
As we demonstrated in our evaluations, these attacks can degrade vehicle safety and reliability, exposing the system to potential risks and leading to non-compliance with emission regulations.
Furthermore, they can impact the data transmission processes for ECU flashing and firmware updates, disrupting their normal execution.
To address this, future standards must incorporate security-oriented enhancements, and vehicle OEMs need to consider appropriate mitigation measures when implementing in-vehicle transport protocols based on the ISO 15765-2 standard.
However, existing countermeasure frameworks such as AUTOSAR SecOC and ISO/SAE 21434 do not fully cover the attack surface addressed in this work. SecOC authenticates PDU payloads at the application layer and therefore cannot detect adversarial manipulation of transport-layer control fields (e.g., FF\_DL, CF\_SN, FC\_FS) or session-override attempts that abort reception below the PDU boundary. ISO/SAE 21434 prescribes a risk-management process but does not mandate transport-layer controls. Consequently, attacks \AID{A4}--\AID{A6} remain effective even in vehicles that deploy SecOC. To address this gap in current automotive security standards, \autoref{tab:attack_mitigation} briefly proposes mitigation strategies for each derived attack.
Furthermore, beyond the transmission of in-vehicle diagnostic data, it is necessary to validate the derived attacks and evaluate their extended impact in broader contexts, such as general vehicle control.

\begin{table}[t]
\centering
\caption{Brief mitigation strategies for attack scenarios based on the ISO 15765-2 standard.}
\label{tab:attack_mitigation}
\vspace{5pt}
\setlength{\extrarowheight}{0pt}
\renewcommand{\arraystretch}{1.5}
\small
\renewcommand{\tabularxcolumn}[1]{m{#1}}

\begin{tabularx}{\columnwidth}{>{\centering\arraybackslash}m{1.4cm} | >{\raggedright\arraybackslash\setlength{\leftskip}{0.25cm}}X}
    \noalign{\hrule height 1.2pt}
    \textbf{Attack} & \multicolumn{1}{>{\centering\arraybackslash}X}{\textbf{Mitigation Strategy}} \\
    \hline 
    
    \AID{A1} & Stateful tracking to prevent inconsistencies between FF\_DL and FC parameters \\ \hline
    \AID{A2} & Validation of FC frame counting and state synchronization between sender and receiver \\ \hline
    \AID{A3} & Validation of FC frame counting and state synchronization between sender and receiver \\ \hline
    \AID{A4} & Service-based allow-listing for escape sequences in high-volume diagnostics \\ \hline
    \AID{A5} & Immediate discard of discontinuous CF\_SN frames, with logging and IDS hooking against sophisticated attacks \\ \hline
    \AID{A6} & FIFO queue-based sequential management for reception sessions \\ \hline
    \AID{A7} & Immediate discard of frames with undefined FC\_FS values \\ \hline
    \AID{A8} & Dynamic timeout management based on CAN bus load \\ 
    \noalign{\hrule height 1.2pt}
\end{tabularx}
\end{table}
\begin{credits}
\subsubsection{\ackname} 
This work was supported by the National Research Foundation of Korea (NRF) grant funded by the Korea government (MSIT) (No. RS-2024-00359621).
\end{credits}

\bibliographystyle{splncs04}
\bibliography{text/Reference}
\end{document}